\documentclass[a4paper]{article}

\usepackage{INTERSPEECH2020}
\usepackage{amsmath,graphicx,cite,multirow}

\title{Light Convolutional Neural Network with Feature Genuinization\\ for Detection of Synthetic Speech Attacks}
\name{Zhenzong Wu$^1$, Rohan Kumar Das$^{1,*}$, Jichen Yang$^{1,*}$ and Haizhou Li$^{1,2}$}
\address{
  $^1$Department of Electrical and Computer Engineering, National University of Singapore, Singapore\\
  $^2$Kriston AI Lab, China}
\email{wuzhenzong@u.nus.edu.sg, \{rohankd, eleyji, haizhou.li\}@nus.edu.sg}

\begin{document}

\maketitle
\begin{abstract}

Modern text-to-speech (TTS) and voice conversion (VC) systems produce natural sounding speech that questions the security of automatic speaker verification (ASV). This makes detection of such synthetic speech very important to safeguard ASV systems from unauthorized access. Most of the existing spoofing countermeasures perform well when the nature of the attacks is made known to the system during training. However, their performance degrades in face of unseen nature of attacks. In comparison to the synthetic speech created by a wide range of TTS and VC methods, genuine speech has a more consistent distribution. We believe that the difference between the distribution of synthetic and genuine speech is an important discriminative feature between the two classes. In this regard, we propose a novel method referred to as feature genuinization that learns a transformer with convolutional neural network (CNN) using the characteristics of only genuine speech. We then use this genuinization transformer with a light CNN classifier. The ASVspoof 2019 logical access corpus is used to evaluate the proposed method. The studies show that the proposed feature genuinization based LCNN system outperforms other state-of-the-art spoofing countermeasures, depicting its effectiveness for detection of synthetic speech attacks.  

\end{abstract}

\vspace{2mm}


\section{Introduction}

In the recent years, automatic speaker verification (ASV) systems are deployed in different real-world applications~\cite{sv_debut,Das2016,SpeechMarker}. These systems are exposed to spoofing attacks for unauthorized access, hence detection of such attacks attracts much attention~\cite{Li2016_spoof_TD,bib:Attacker_overview2020}. Various spoofing attacks are broadly classified into replay, impersonation, voice conversion (VC) and text-to-speech synthesis (TTS) attacks~\cite{spoof_review}. The latest progress in VC and TTS systems can produce perceptually natural sounding speech, which poises a threat to fool the ASV systems~\cite{Lorenzo-Trueba2018,Kinnunen2018_vc,Obama2018}.  

\footnotetext{*Corresponding Author}

The research on spoofing countermeasures grew in the last decade since the inception of ASVspoof\footnote{http://www.asvspoof.org/} challenge series. The challenge provided a platform to the researchers across different domains to explore fake speech detection using a common benchmarked corpus~\cite{spoof_is2013,ASVspoof_journal}. Its recent edition ASVspoof 2019 is devoted to detection of both synthetic and replay speech with two subtasks~\cite{ASVsppof2019_paper}. The logical access track focuses on detection of synthetic speech created using state-of-the-art VC and TTS systems, which is the focus of this paper.


The explorations on spoofing attack detection cover two directions from the perspective of a detection task. The spoofing countermeasures either focus on novel front-end features or effective classifiers. 
Some of the former studies focused on the importance of robust features such as cochlear filter cepstral coefficient and instantaneous frequency (CFCCIF)~\cite{PatelINTERSPEECH2015}, linear frequency cepstral coefficients (LFCC), subband spectral flux coefficients and spectral centroid frequency coefficients~\cite{ASVspoof2015_lfcc}. Later, the long-term constant-Q transform (CQT) based constant-Q cepstral coefficients (CQCC) proved to be one of the strong front-ends for synthetic speech detection~\cite{CQCC_odyssey2016}. The recent explorations with features derived from CQT are also found to effective for spoof detection~\cite{taslp_cmc,tifs_subband,elsevier_icqcc}.

With the advent of deep learning methods, robust classifiers are investigated for detection of spoofing attacks. Some of these include end-to-end systems with light convolutional neural networks (LCNN)~\cite{galinaITNTEERSPEECH2019,Yang2019}, squeeze excitation and residual networks~\cite{chengilaiITNTEERSPEECH2019,Alam_ASRU}. The end-to-end systems have much difference with the works that focus on novel features. The former are data driven deep learning methods, while the latter emphasize on hand-crafted feature, which require prior knowledge. Further, we note that the same neural network based system can perform differently for a range of features~\cite{galinaITNTEERSPEECH2019}. Therefore, a robust spoofing countermeasure is required to have a strong feature extractor that captures the discriminative artifacts along with an effective classifier. 

The synthetic speech attacks can be created with a wide range of TTS and VC algorithms~\cite{spoof_review}. In general, spoofing countermeasures do not handle synthetic speech from unseen sources because of lack of generalization ability~\cite{Gen_CM_rkd}. We note that  genuine examples have a comparatively lower variance than synthetic speech. We believe that the consistent characteristics of genuine speech set genuine speech apart from a  variety of different synthetic speech. A recent study using temporal domain information shows that spoofing detection can be improved by modifying the probability mass function of spoofed speech close to that of the genuine speech~\cite{genuinization2019}. This process is termed as genuinization, which is found to be effective when applied to both train and test examples for synthetic speech detection.  


In a similar direction, we hypothesize that, if we are able to derive a model that fits well the distribution of the genuine speech, such a model will take genuine speech as the input and generate genuine speech as the output following the same distribution of the genuine speech. However, when the model takes spoof speech as input, it will generate very different output, that amplifies the difference to genuine speech. With this hypothesis, we propose to derive a model using the genuine speech features with convolutional neural network (CNN) that is referred to as genuinization transformer. Further, the process is referred to as feature genuinization as a given feature representation is projected on a domain learned using the genuine features. The genuinization transformer is then used together with an LCNN system for detection of synthetic speech attacks.


The rest of the paper is organized as follows. Section~\ref{secii} introduces the details of proposed feature genuinization. Section~\ref{seciii} describes the feature genuinization based LCNN system for detection of spoofing attacks. The experiments and their results with discussion are reported in Section~\ref{seciv} and Section~\ref{secv}, respectively. Finally, the paper is concluded in Section~\ref{conc}.

\section{Feature Genuinization}
\label{secii}

\begin{figure}[t]
  \centering
  \includegraphics[width=\linewidth, height=6cm]{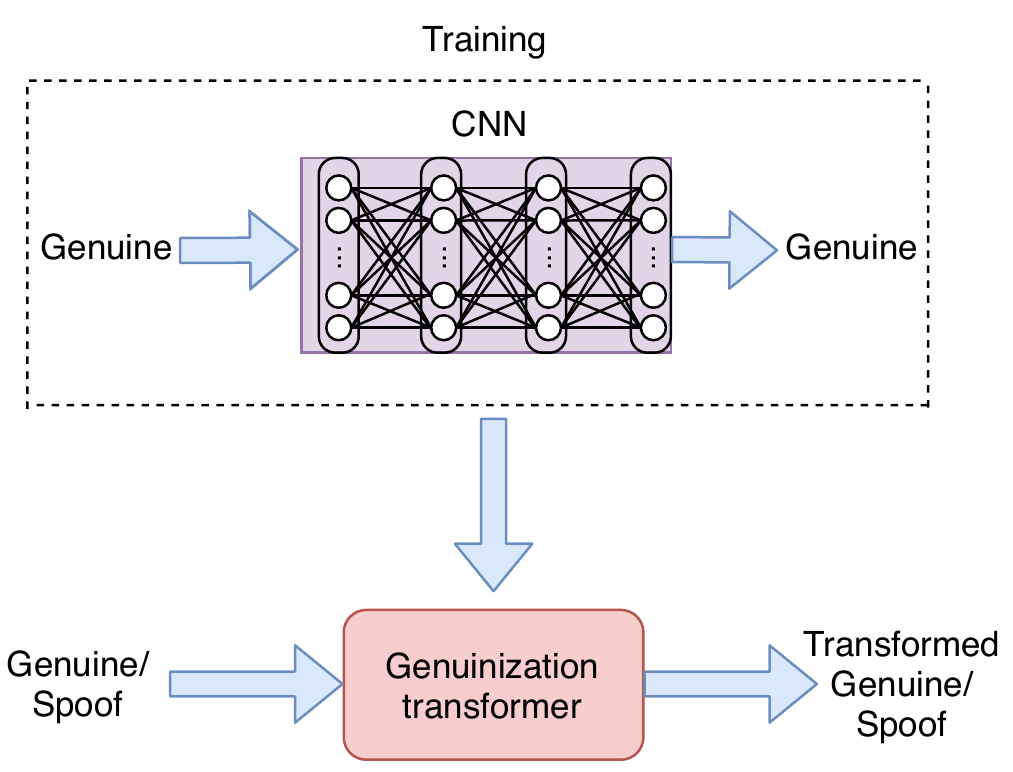}
  \caption{The block diagram of feature genuinization process.}
    \label{gentransformer1}
  \vspace{-2mm}
\end{figure}

\begin{figure}[ht]
  \centering
  \includegraphics[width=0.8\linewidth, height=12.8cm]{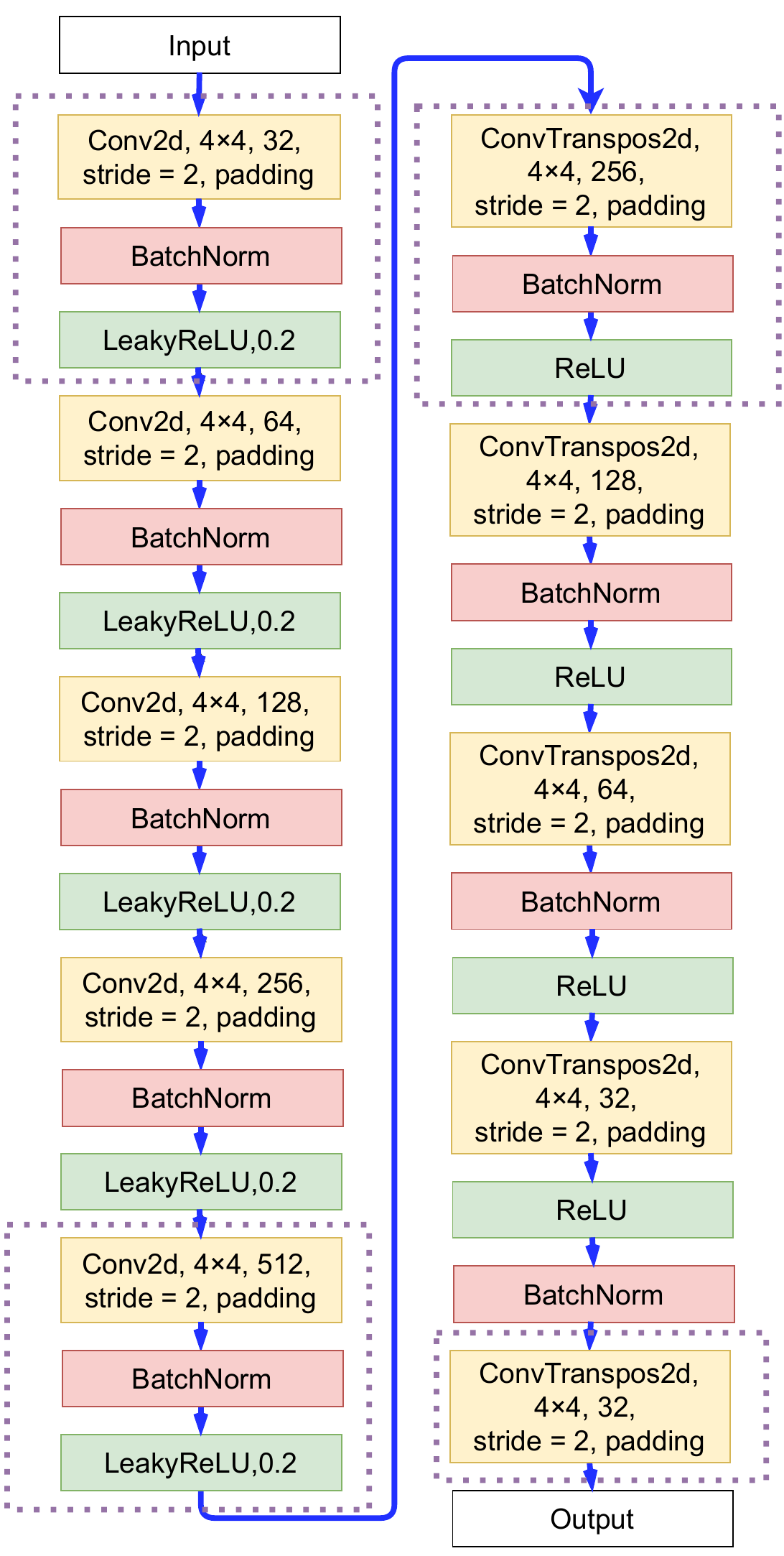}
  \caption{The architecture of genuinization transformer.}
  \label{gentransformer2}
  \vspace{-3mm}
\end{figure}

\begin{figure}[ht]
  \centering
  \includegraphics[width=4cm, height=6cm]{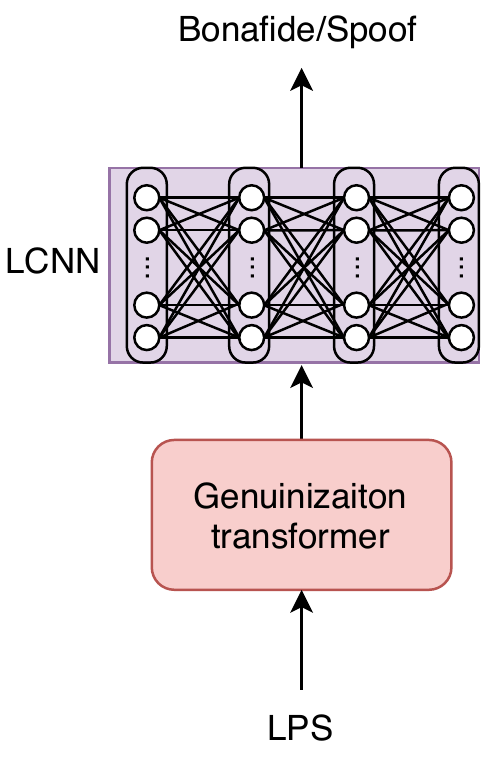}
  \vspace{-2mm}
  \caption{Block diagram of the proposed feature genuinization based LCNN system.}
  \label{LCNN}
  \vspace{-4mm}
\end{figure}


We aim to learn a transformer that does not change the characteristics of genuine speech features, whereas it projects spoof speech to a different output, maximizing the difference between genuine and spoof speech. Figure~\ref{gentransformer1} shows the block diagram of the proposed feature genuinization process. It can be observed that there are two stages of the process. The first stage basically focuses on training a feature genuinization transformer using the characteristics features derived from only genuine speech. During the second stage, this trained feature genuinization transformer is used to convert any given features that enhances the discrimination of genuine and spoof speech.




The CNN based architectures have shown their effectiveness in the field of anti-spoofing research~\cite{zhangchunleiICASSP2016}. In this regard, we use CNN for training the genuinization transformer as shown in Figure~\ref{gentransformer1}. The detailed architecture of the CNN used in this framework can be seen from Figure~\ref{gentransformer2}. It can be observed that the functionality of the proposed genuinization transformer is similar to that of an autoencoder. However, the output of genuinization transformer is considered as the final transformer result. In addition, we apply a full convolutional layer and therefore, there is no fully connected layers in the transformer. This can thereby force the network to focus on the temporal correlation between the input signal and the whole stratification process. Further, it reduces the number of training parameters, which significantly results in less training period.



A study in~\cite{DCGAN2015} shows that it is a good practice to use strided convolution rather than pooling to downsample as it allows the network to learn its own pooling function. Therefore, we use this method during the training of genuinization transformer. In addition, batchNorm2d and leaky rectified linear unit (ReLU) activation function are used in the training because they can promote healthy gradient flow, which is critical for the learning process.


The architecture of proposed genuinization transformer shown in Figure~\ref{gentransformer2} consists of two functionalities: encoding and decoding. During the encoding phase, the input signal is compressed through a number of strided convolutional layers, and then the convolution result is obtained by leaky ReLU. 
In the decoding phase, the encoding process is reversed by deconvolution, and then by ReLU. In this way, the transformer works as an autoencoder that learns the characteristics of genuine speech~\cite{Deepautoencoder}. As a result of this, it amplifies the discrimination of genuine and spoof speech in the transformed domain.


Once the genuinization transformer is trained, it can be used to transform any given genuine/spoofed features to a transformed domain that is learned using the only genuine feature characteristics. This novel way of transforming the feature is referred as feature genuinization as mentioned earlier. Next, we discuss about the LCNN system using the feature generalization for detection of spoofing attacks.

\section{LCNN with Feature Genuinization}
\label{seciii}

Various deep learning systems have shown their effectiveness for spoofing attack detection~\cite{galinaITNTEERSPEECH2019,chengilaiITNTEERSPEECH2019,Jung2019,Gomez-Alanis2019,Alam_ASRU}. Therefore, we plan to use the proposed feature genuinization with a deep learning system. The LCNN is one of the strongest systems that has proven to be useful for its compactness and efficacy for anti-spoofing~\cite{wu2018light,galinaITNTEERSPEECH2019}. In this work, we use LCNN based system with the transformed features obtained using genuinization transformer.


Figure~\ref{LCNN} shows the block diagram of the proposed feature genuinization based LCNN system. We consider log power spectrum (LPS) of a given speech as the input feature to the genuinization transformer. It transforms the given input LPS to a genuinized feature, which is an input to the LCNN. During training, the training data and their corresponding label information is fed to the LCNN system. Once the training is completed, the detection result for a given input to the system can be obtained to identify the spoofing attacks.


We used Max-Feature-Map (MFM) activation function instead of commonly used ReLU function for the LCNN system similar to that in~\cite{galinaITNTEERSPEECH2019}. The main advantage of MFM is that it can learn compact features instead of sparse high-dimensional ones like ReLU. Further, MFM resorts to max function to suppress the activations of a small number of neurons so that MFM based CNN models are light and robust. Therefore, these are applied to reduce the dimensionality of the output and obtain more discriminative feature maps.



\section{Experiments}
\label{seciv}

In this section, we discuss the database and experimental setup for the studies.

\subsection{Database}

\footnotetext[2]{https://datashare.is.ed.ac.uk/handle/10283/3336}
\footnotetext[3]{http://dx.doi.org/10.7488/ds/1994}
\footnotetext[4]{https://pytorch.org}

We consider the ASVspoof 2019 logical access corpus\footnotemark[2] for the studies of synthetic speech detection in this work~\cite{ASVspoof2019_plan,ASVsppof2019_paper}. The corpus has three partitions, which are train, development and evaluation set. The genuine examples of the ASVspoof 2019 corpus are part of VCTK\footnotemark[3] database, which is a standard corpus for speech synthesis. It contains 107 speakers data that includes 46 male and 61 female speakers. It is to be noted that there is no overlap of speakers across different subsets. The synthetic speech attacks for the development set are created with two VC and four TTS state-of-the-art methods. However, the spoofed examples of evaluation set are derived from unseen methods. The ASVspoof 2019 uses an ASV-centric metric given by tandem detection cost function (t-DCF) as the primary metric and equal error rate (EER) as a secondary metric for benchmarking the systems~\cite{ASVspoof2019_plan,Kinnunen2018_TDCF}. We considered the scores of ASV system given along with the ASVspoof 2019 logical access corpus to combine with that from spoofing countermeasure system for computation of t-DCF measure. Table~\ref{tab:la2019} presents a summary of the ASVspoof 2019 logical access corpus.

 \begin{table}[!t]
 \begin{center}
 \caption{Summary of ASVspoof 2019 logical access corpus.}
 \label{tab:la2019}
\resizebox{8.1cm}{!}{
 \begin{tabular}{|c|c|c|c|c|}
 \hline {\bf{Subset}}       &{\bf{\#Male}} &{\bf{\#Female}}  &{\bf{\#Bonafide}} &{\bf{\#Spoofed}} \\
 \hline
 \hline Train     & 8   & 12     & 2,580    & 22,800 \\
 \hline Development  & 4   & 6     & 2,548    & 22,296 \\
 \hline Evaluation   & 21  & 27     & 7,355    & 63,882 \\
 \hline
 \end{tabular}
 }
 \end{center}
 \vspace{-8mm}
 \end{table}

\subsection{Experimental Setup}

The long-term CQT based features are found to capture useful artifacts for spoofing attack detection~\cite{rkd_is2019}. Therefore, we use LPS derived from CQT as the input feature for the studies. The parameters for CQT computation are set based on following those in~\cite{CQCC_odyssey2016}. The number of octaves and frequency bins in every octaves are set at 9 and 96, respectively. In addition, the static dimension of LPS is 863. For LPS extraction from CQT, the length of every file is set as 256 frames by either padding and cropping. In particular, the examples with frame-length over than 256 frames are truncated, while the examples with frame-length less than 256 frames are filled with the last frame value. Thus, the we have an input feature of 863$\times$256 for every example.

During training of the LCNN system, an additional batch normalization step is used after max pooling layer to increase the stability and convergence speed. As such models are prone to overfitting, we consider dropout and weight decay methods to avoid such issue. The dropout is used for fully connected layers with the ratio 0.4 and the weight decay is set to $2\times10^{-4}$. In addition, the parameters like number
of layers and nodes are optimized on the development set. The proposed feature genuinization based LCNN system is implemented using PyTorch\footnotemark[4] toolkit. 





\section{Results and Discussion}
\label{secv}


The proposed system is a pipeline with a feature genuinization followed by LCNN. We compare the proposed system with LCNN baseline without feature genuinization. Further, we also consider the two baseline spoofing countermeasures of ASVspoof 2019 challenge. They are based on CQCC and LFCC features with Gaussian mixture model (GMM) classifier~\cite{ASVspoof2019_plan,ASVsppof2019_paper}.


Table~\ref{tab:la2019dev} shows the results of proposed feature genuinization based LCNN system, that we refer as FG-LCNN, on ASVspoof 2019 logical access corpus and its comparison to the baseline systems. We observe that introducing feature genuinization module in the baseline LCNN system improves the detection of spoofing attacks. While the results on the development set are close, the improvement from the proposed system is evident from the results on the evaluation set, which contains more challenging spoofing attacks of unseen nature. This confirms our hypothesis to use a feature genuinization model exploiting the characteristics of genuine speech. Further, we find that the performance of the proposed system is much better than the two ASVspoof 2019 challenge baselines. 



\begin{table}[!t]
\begin{center}
\caption{Performance of proposed feature genuinization based LCNN (FG-LCNN) and its comparison to baseline systems on ASVspoof 2019 logical access corpus.}
\label{tab:la2019dev}
\resizebox{8.1cm}{!}{
\begin{tabular}{|c|c|c|c|c|}
\hline
\multirow{2}{*}{\bf{System}} & \multicolumn{2}{|c|}{\bf{Development Set}}& \multicolumn{2}{|c|}{\bf{Evaluation Set}}\\
\cline{2-5}
     &{\bf{t-DCF}}           &{\bf{EER (\%)}}      &{\bf{t-DCF}}           &{\bf{EER (\%)}}   \\ \hline
\hline Baseline: LCNN         &0.002           &0.080     &0.111           &4.448 \\
{\bf FG-LCNN}     &{\bf 0.000}           &{\bf 0.002}     &{\bf 0.102}           &{\bf 4.070}  \\
\hline
\hline
\multicolumn{5}{|c|}{\bf{ASVspoof 2019 Baseline~\cite{ASVsppof2019_paper}}}\\
\hline
CQCC-GMM & 0.0123& 0.43  &0.2366 & 9.57 \\
LFCC-GMM &0.0663 & 2.71  & 0.2116& 8.09\\
\hline
\end{tabular}
}
\end{center}
\end{table}

\begin{table}[!t]
\begin{center}
\caption{Performance of proposed feature genuinization based LCNN (FG-LCNN) and its comparison to feature spoofing based LCNN (FS-LCNN) contrast system on ASVspoof 2019 logical access corpus evaluation set.}
\label{tab:contrast}
\begin{tabular}{|c|c|c|}
\hline
{\bf{System}}&{\bf{t-DCF}}           &{\bf{EER (\%)}}   \\ \hline
\hline Baseline: LCNN          &0.111           &4.448 \\
\hline
\hline
{\bf Prposed: FG-LCNN}        &{\bf 0.102}           &{\bf 4.070}  \\
Contrast: FS-LCNN        &0.138           &4.860  \\
\hline
\end{tabular}
\end{center}
\vspace{-2mm}
\end{table}

We further perform a justification experiment for validation of our proposed method. The idea behind feature genuinization process is based on the assumption that the genuine speech examples are considered to be less varied than the synthetic speech attacks created using a wide range of methods. We perform a contrast experiment, where we learn a transformation model using CNN by only considering the spoofed speech features. We refer this process as feature spoofing and the model as spoofing transformer, similar to the case of our proposed method. This spoofing transformer is then used to transform any given feature of genuine or spoofed speech to another domain, which is then used in the LCNN system pipeline, that we call FS-LCNN. The rest of the experimental setup remains the same to that our proposed method.

Table~\ref{tab:contrast} shows the performance comparison of the FS-LCNN contrast system with our proposed FG-LCNN and the baseline LCNN system. We consider the results of evaluation set for the comparison as the results of development set can show very accurate detection of synthetic speech attacks. We find that the FS-LCNN contrast system does not perform better than our proposed FG-LCNN system, but rather degrades from the baseline LCNN system. This further strengthens our proposed idea of using feature genuinization process with LCNN system for detection of spoofing attacks. 

We are now interested in comparing the proposed system to various single system based results available of ASVspoof 2019 logical access corpus. In this regard, we consider some of the well performing front-end as well as back-ends that have shown their effectiveness for spoofing attack detection in ASVspoof 2019 challenge. Some of the those front-ends  are zero time windowing cepstral coefficients (ZTWCC), single frequency cepstral coefficients (SFCC) and instantaneous frequency cepstral coefficients (IFCC) that are implemented with GMM based classifier~\cite{knrkrajuITNTEERSPEECH2019}. Further, deep learning based classifiers such as deep neural network (DNN), ResNet and LCNN are used for detection of spoofing attacks using front-ends like mel frequency cepstral coefficient (MFCC), constant-Q statistics-plus-principal information coefficients (CQSPIC), CQCC, LFCC, LPS of discrete Fourier transform (DFT) and fast Fourier transform (FFT) in ASVspoof 2019 challenge~\cite{rkd_ASRU2019,moustafaITNTEERSPEECH2019,galinaITNTEERSPEECH2019,applied_acoustics}. We report the respective system results from their published works for the comparison on the evaluation set of ASVspoof 2019 logical access corpus.

\begin{table}[!t]
 \begin{center}
 \caption{Performance comparison of the proposed feature genuinization based LCNN system to some known single systems on ASVspoof 2019 logical access evaluation set.}
 \label{la2019comsys}
 \begin{tabular}{|c|c|c|}
 \hline {\bf{System }}                                                                    &{\bf{t-DCF}}                     &{\bf{EER (\%)}}    \\  \hline
  \hline ZTWCC-GMM~\cite{knrkrajuITNTEERSPEECH2019}                                 &0.141                     &6.13         \\
 \hline IFCC-GMM~\cite{knrkrajuITNTEERSPEECH2019}                                  &0.357                     &15.59         \\
 \hline SFFCC-GMM~\cite{knrkrajuITNTEERSPEECH2019}                                 &0.323                     &13.97         \\
 \hline CQCC-DNN~\cite{rkd_ASRU2019}                                         &0.308                     &12.79        \\
 \hline LFCC-DNN~\cite{rkd_ASRU2019}                                         &0.234                     &9.65         \\ 
 \hline MFCC-ResNet~\cite{moustafaITNTEERSPEECH2019}                               &0.204                     &9.33         \\
 \hline LPS-DFT-ResNet~\cite{moustafaITNTEERSPEECH2019}                               &0.274                     &9.68         \\
 \hline CQCC-ResNet~\cite{moustafaITNTEERSPEECH2019}                               &0.217                     &7.69         \\
 \hline CQSPIC-DNN~\cite{rkd_ASRU2019}                                           &0.183                     &7.81         \\
 \hline CQSPIC-GMM~\cite{rkd_ASRU2019}                                           &0.164                     &7.74         \\
 \hline LFCC-LCNN~\cite{galinaITNTEERSPEECH2019}                                   &0.100                     &5.06         \\
 \hline LPS-FFT-LCNN~\cite{galinaITNTEERSPEECH2019}                                    &0.103                     &4.53         \\ \hline
 \hline {\bf {Proposed: FG-LCNN}}                                             &{\bf {0.102}}   &{\bf {4.07}}         \\
 \hline
 \end{tabular}
 \end{center}
 \vspace{-2mm}
 \end{table}


Table~\ref{la2019comsys} reports the performance comparison of the proposed FG-LCNN system to some of the single systems reported in ASVspoof 2019 challenge discussed above. It is observed that the LCNN based systems represent the best performing single system, that justifies its use as the baseline LCNN in this work. Further, the effectiveness of proposed feature genuinization is evident on using it with the LCNN system, which outperforms other reported single systems in terms of EER on ASVspoof 2019 logical access corpus.


\section{Conclusion}
\label{conc}



This work proposes a novel feature genuinization based LCNN system for detection of synthetic speech attacks. The characteristics of genuine speech are exploited to learn a model using CNN. It transforms a genuine feature distribution more close to that of the genuine speech, whereas leads to a different output for features of spoof speech, thereby maximizing their difference. The transformed features are then used with an LCNN system. The studies conducted on ASVspoof 2019 logical access corpus show the effectiveness of the feature genuinization based LCNN system for detecting synthetic speech attacks. The comparison of the proposed system to various state-of-the-art spoofing countermeasures showcases it as one of the strong single anti-spoofing system. The future work will focus on extending the studies to replay attack detection.

\section{Acknowledgements}

This research work is partially supported by Programmatic Grant No. A1687b0033 from the Singapore Government's Research, Innovation and Enterprise 2020 plan (Advanced Manufacturing and Engineering domain), Human-Robot Interaction Phase 1 (Grant No. 192 25 00054) by the National Research Foundation, Prime Minister's Office, Singapore under the National Robotics Programme. This work is also part of a collaboration with Kriston AI Lab, China in 2019.


\bibliographystyle{IEEEtran}

\bibliography{MyReferences_new}

\end{document}